\documentclass[aps,prd,twocolumn,showpacs,nofootinbib]{revtex4-1}

\usepackage[latin1]{inputenc}

\usepackage{latexsym}
\usepackage{amsbsy}
\usepackage{mathrsfs}
\usepackage{color}

\usepackage{amsmath,amssymb,calc,amsfonts}
\usepackage{latexsym}
\usepackage{hyperref}
\usepackage{wasysym}

\usepackage{graphicx,calc,epsfig}
\def\ut#1{\rlap{\lower1ex\hbox{$\sim$}}#1{}}

\newcommand{\be}{\nopagebreak[3]\begin{equation}}
\newcommand{\ee}{\end{equation}}
\newcommand{\ba}{\nopagebreak[3]\begin{eqnarray}}
\newcommand{\ea}{\end{eqnarray}}
\DeclareFontFamily{U}{rsfs}{}         
\DeclareFontShape{U}{rsfs}{m}{n}{<5> rsfs5 <6><7> rsfs7          %
  <8><9><10><10.95><12><14.4><17.28><20.74><24.88> rsfs10}{}     %
\DeclareMathAlphabet{\mathfs}{U}{rsfs}{m}{n}                     %
\def\pb#1{\rlap{\lower1.5ex\hbox{$\longleftarrow$}}{#1}}
\def\dpb#1{\rlap{\lower1.5ex\hbox{$\Longleftarrow$}}{#1}}
\def\spb#1{\rlap{\lower1.5ex\hbox{$\leftarrow$}}{#1}}
\def\sdpb#1{\rlap{\lower1.5ex\hbox{$\Leftarrow$}}{#1}}

\definecolor{blue}{rgb}{0,0,1}
\definecolor{green}{rgb}{0,0.6,0.5}
\definecolor{red}{rgb}{1,0,0}
\definecolor{vio}{rgb}{0.6,0,1}
\definecolor{ama}{rgb}{1,1,0}

\begin{document}

\title{
Intrinsic angular momentum for radiating spacetimes \\
which agrees with the Komar integral in the axisymmetric case
}

\date{\today}

\author{Emanuel Gallo and Osvaldo M. Moreschi}

\affiliation{FaMAF, Universidad Nacional de Córdoba \\
Instituto de Física Enrique Gaviola (IFEG), CONICET \\
Ciudad Universitaria, (5000) C\'ordoba, Argentina. }

\begin{abstract}

Here, we present a new definition of {intrinsic angular momentum} at future null infinity,
based on the charge-integral approach. This definition is suitable for the general case of 
radiating spacetimes without symmetries, which does not suffer from supertranslations 
ambiguities. In the case of axial symmetry this new definition agrees with 
the Komar integral.

\end{abstract}


\maketitle

\section{Introduction}

The subject of physical quantities, as \emph{momentum} and \emph{intrinsic angular momentum},
is always related to the notion of symmetries of the physical system, to which
they refer to.
Thus, in the framework of special relativity, we have at our disposal the definitions
of total momentum and total angular momentum, based on the existence of
the 10 Killing symmetries of the spacetime.
When, considering the analogous situation for an isolated system, in the framework 
of general relativity things are more complicated.
One is faced with the difficulty, 
in the neighborhood of future null infinity ($\mathcal{I}^+$),
the asymptotic symmetries generate an \emph{infinite dimensional} group; namely, 
the Bondi-Metzner-Sachs (BMS)\cite{Bondi62,Sachs62,Sachs62b} one.
This is because the existence of gravitational radiation affects the curvature
of the spacetime, even in the asymptotic regime.

In the study of \emph{asymptotically flat} spacetimes,
it is tempting to express the asymptotic structure in terms of a decomposition of the 
metric over a {flat} one, namely:
\begin{equation}\label{eq:asymp1}
 g_{ab} = \eta_{ab} + h_{ab}
;
\end{equation}
where {$\eta_{ab}$ is a flat} metric and $h_{ab}$ the tensor in which all the physical information
is encoded. But the problem is that there are as many flat metrics as there are  
proper BMS\cite{Sachs62,Moreschi86}
supertranslation generators. 
We should probably note that this difficulty is rooted in the existence of gravitational
radiation, which reaches future null infinity; 
where total physical quantities are calculated.
On the contrary, in the case of a stationary system, which therefore has no gravitational
radiation content, one can single out a unique flat background metric $\eta_{ab}$,
which one can use in (\ref{eq:asymp1}).

The situation today is that there are numerous references 
for definitions of angular momentum 
at future null infinity, for general radiating spacetime;
most of them suffer from the so called problem of supertranslation 
ambiguities\cite{Bramson75,Prior77,Winicour80,Geroch81,Dray84};
but the physics community has
not yet embraced a standard of it (for a review and references on the subject of 
energy-momentum and angular momentum in general relativity
see \cite{Szabados04}).
This is in spite of the fact that in \cite{Moreschi04}
a definition of \emph{intrinsic angular momentum} was presented,
free from supertranslation ambiguities, satisfying a set of appropriate physical conditions.
However, in that reference, it was not resolved the relation between that definition
and the Komar integral,  in the case of the existence of an axial symmetry.
Although the presence of a rotational Killing vector is not the usual situation;
its study gives important clues on invariant definitions of angular momentum.
Since under these circumstances, the Komar integral becomes an important tool, due to its
conservation properties, in this article we tackle this problem presenting
a new definition of intrinsic angular momentum, free from supertranslation ambiguities,
and which agrees with the Komar integral,
in the case of the particular situation of the existence of an axial symmetry.
Therefore, the main improvement of this work over 
\cite{Moreschi04}, which has already provided with a definition of 
intrinsic angular momentum free of supertranslation ambiguities,
is to answer our question\cite{Moreschi04} and other criticisms\cite{Kozameh:2012hx}
about the relation of this approach to the Komar integral.

It is probably worthwhile to remark that most of the numerous definitions of angular
momentum at future null infinite do not tackle the problem of \emph{intrinsic}
quantities.

In our work we will use extensively a definition of rest frames\cite{Moreschi88,Moreschi98,Dain00'}, 
which is described below;
along with its relation to definitions of \emph{center-of-mass}
and intrinsic angular momentum. 
Using these definitions we can further select a \emph{unique} {timelike orderly family
of sections} that have the information for {center-of-mass} and 
{intrinsic angular momentum}\cite{Moreschi04}.

Therefore, for each point at future null infinity, we have\cite{Moreschi04} {a way to single out
a unique decomposition of the metric in the form (\ref{eq:asymp1}), with an
appropriately selected flat background $\eta$}.

\begin{figure}[h]
\includegraphics[clip,width=0.45\textwidth]{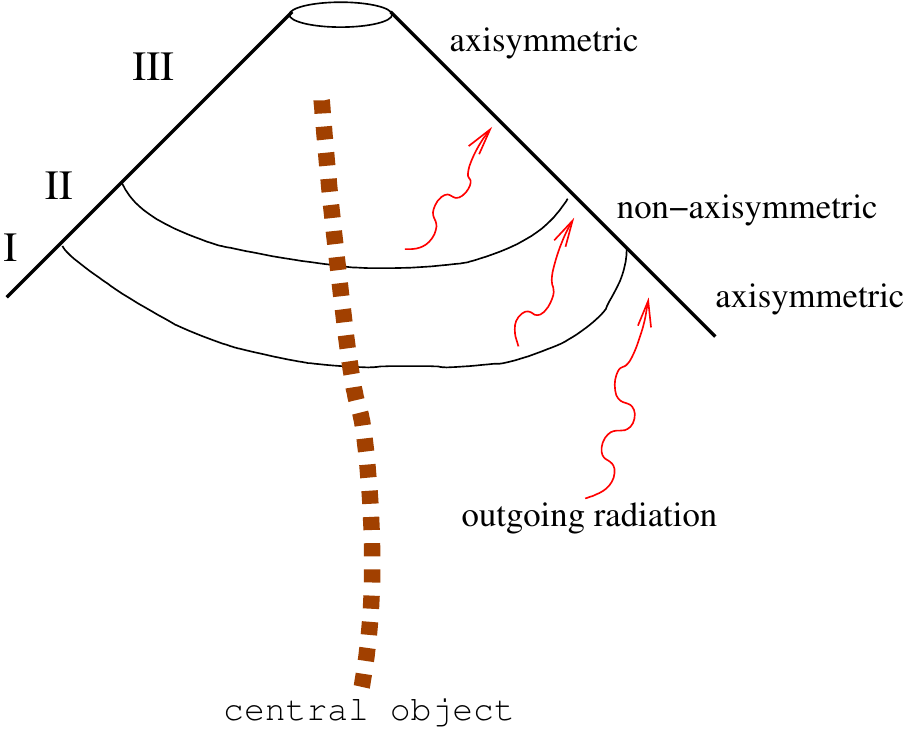}
\caption{Let us consider a spacetime such that its asymptotic regions has three
different stages as depicted in this figure.
}
\label{fig:centerofmass0}
\end{figure}

\begin{figure}[h]
\includegraphics[clip,width=0.4\textwidth]{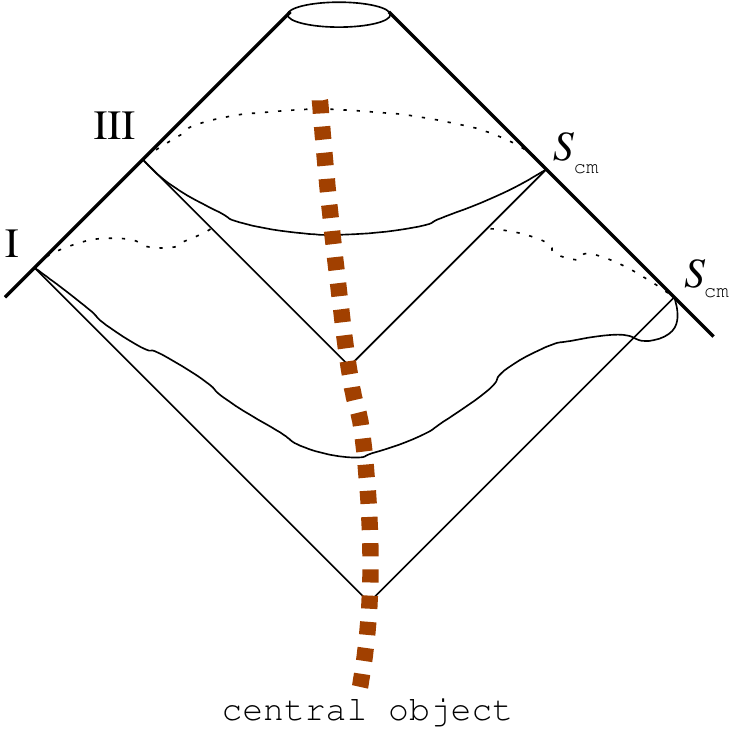} 
\caption{Any {\bf intrinsic definition of center-of-mass}
will {necessarily involve supertranslation} among different instances of the
dynamical system.
}
\label{fig:centerofmass1}
\end{figure}

In order to gain perspective of our work
let us consider a radiating asymptotically 
flat spacetime in which one can distinguish three stages.
The first stage is when the asymptotic region possesses an axial symmetry,
the second stage is when no symmetry is found; and a third stage in which one finds
another axial symmetry, which does not coincide with that of the first stage.
The situation is depicted in figure \ref{fig:centerofmass0}.
Here, we present new definitions of center-of-mass and intrinsic angular momentum 
which are suitable for the general case of 
radiating spacetimes, 
that do not suffer from supertranslation 
ambiguities and which
in the first and third stages, where two different rotational Killing vectors exist,
coincide with the Komar integral.
To our knowledge, these are the only definitions of
center-of-mass and intrinsic angular momentum which possess these properties.
It is important to emphasize that the corresponding center-of-mass in the 
first and third stages would, in general, involve
a supertranslation (See figure \ref{fig:centerofmass1}). 

Our work is based on the charge integral approach. In the past, Penrose\cite{Penrose82} 
has used the notion of {charge integrals of the Riemann}
tensor in his study of {quasilocal mass and angular momentum}. 
We\cite{Moreschi86} have used the same approach in our work on 
{total angular momentum}, with a global choice of reference frame. 
Later, in \cite{Moreschi04}, we used this concept in our 
construction of a
definition of {intrinsic angular momentum} {free of supertranslation ambiguities}.
One advantage of the approach of charges integral is that for each generator of 
asymptotic symmetry one have an associated physical quantity; in particular, the same expression provides 
not only a definition of angular momentum but also of total linear momentum. 
Furthermore, by construction, the factor of two anomaly problem,
found in the Komar integrals\cite{Winicour80}, is absent in this formalism. 

We are presenting, for the first time a definition of intrinsic angular momentum
in general relativity which is free from supertranslation ambiguities,
valid for the most general class of isolated systems, and it agrees with
the Komar integral in the presence of rotational Killing vectors.
We think that this work will be of interest for a variety of readers;
some of them might not be experts in the subject, but the definition 
might be relevant for their work.
For this reason we try to present it in a form as self-contained as possible,
so that some expert readers are advised to overlook a sections with
standard definitions.

The paper is organized as follows. 
In Section II, we present a brief review of the 
concepts of rest frames, supermomentum and nice sections 
that readers already acquainted with these notions may omit.
In Section III the Komar integral is expressed in terms of spinorial quantities 
using the Geroch-Held-Penrose(GHP)\cite{Geroch73} formalism.
In Section IV and V, we discuss the charge-integral approach, and finally in the last section 
we present the new construction and definition of intrinsic angular momentum valid for 
general asymptotically flat spacetimes.


\section{`Nice sections' as a tool for the supertranslation problem}

\subsection{Too many rest frames}

As was mentioned in the introduction, in spacetimes which are asymptotically flat at future null infinity, 
instead of the Poincare group 
with a finite number of symmetries, we have the BMS group, with an infinite number 
of asymptotic symmetries. 
Although this group contains the subgroup of translations as a normal subgroup 
(which allows us to define geometrically the linear total Bondi momentum), it does not contain 
a subgroup of Lorentz rotations defined in a canonical way. 
Therefore, there is no a priori 
intrinsic way to define a Pauli-Lubanski like vector.
To see some specific examples where these constructions would fail, we refer to
the work \cite{Gallo09}; where it is shown, in particular, that a supertranslated
boosted section in Schwarzschild spacetime, would give a nonzero angular momentum;
which if used in a 
Pauli-Lubanski like vector\cite{Lousto07}, would produce
a quantity that is supertranslation dependent.
However, we can base the analysis in terms of the supermomentum;
which is defined in terms of the infinite 
supertranslation generators and which we use to define the concept of \emph{nice sections}.
The supermomentum is an object with 
infinite components such that the first four of them define the total Bondi momentum. 
Using this 
concept, we can ask whether there are sections at $\mathcal{I}^+$ such that on these sections 
the only nonvanishing component of the supermomentum is
the {first}  one; that is, the timelike component of the Bondi momentum. 
This would allow to give a definition of rest-frames
that we materialize in the concept of nice sections. 
But for this program to really work, one 
should prove, among other things, that this family of sections is actually a four-parameter family. 
Fortunately, it was proved in the past\cite{Moreschi88,Moreschi98,Dain00'} 
 that this program can be successfully carried out. 
With the concept of nice sections at hand, we have at our disposal a tool to 
single out a Lorentz subgroup of the BMS group for each rest frame.
We will now review, some of these concepts.

\subsection{Bondi systems as inertial frames}

The analogues to {inertial frames} of special relativity at future null infinity are
the {Bondi systems}; by which we mean\cite{Moreschi86} 
a {coordinate and tetrad system}; which we describe below.
Let $(M, g_{ab})$ be an asymptotically flat spacetime at null infinity\cite{Penrose63}.
As it was mentioned in the Introduction,
in a vicinity of $\mathcal{I}^+$ one can express the metric in terms
of a flat metric $\eta$ plus a tensor $h$; 
where $h$ goes to zero appropriately\cite{Moreschi87} as one approaches future null infinity. Each Bondi system can be used to build such a flat metric $\eta$;
and any other Bondi system connected to the first by a translation or 
a Lorentz rotation, will determine the same flat metric $\eta$.
However, any other Bondi system connected to the first by a proper supertranslation
will determine a {\bf different} flat metric tensor $\eta$.
This situation is what complicates the discussion of global quantities such as angular
momentum at $\mathcal{I}^+$.

Let us explain what do we mean by Bondi systems.
In the vicinity of $\mathcal{I}^+$, we can construct a coordinate 
system $(u,r,\zeta,\bar\zeta)$ where
$u$ are null hypersurfaces, $r$ is an affine parameter of the null generators $l^a=g^{ab}(du)_b$
of the null hypersurfaces $u=\text{const}$ such that when $r$ goes to infinity, the integral curves of
these null generators intersect $\mathcal{I}^+$, and $(\zeta,\bar\zeta)$ are stereographic coordinates
labelling the null generators of $\mathcal{I}^+$. If this coordinate system is chosen such that 
when $r$ goes to infinity the induced (conformal) intrinsic metric on $\mathcal{I}^+$ $\hat{g}_{ab}=\Omega^2g_{ab}$ with
$\Omega=r^{-1}$ is the standard metric of a unit sphere,
or more precisely, if the metric $\hat{g}|_I$ on $\mathcal{I}^+$ reads
\begin{equation}
\hat{g}|_I=0\cdot du^2-4\frac{d\zeta d\bar\zeta}{(1+\zeta\bar\zeta)^2} \, ;
\end{equation}
then the coordinate system $(u,\zeta,\bar\zeta)$ defines a Bondi system.
One can further require the affine coordinate $r$ to agree with the so called
`luminosity distance'\cite{Moreschi87}; so that the coordinate system has an invariant
extension into the interior of the spacetime.
Associated to this coordinate system, we have a null tetrad $\{l^a,n^a,m^a,\bar{m}^a\}$,
in the vicinity of $\mathcal{I}^+$,
 where
$n^a$ is a null vector such that $l^an_a=1$, with $m^a\bar{m}_a=-1$
and all other possible products vanishing; 
$m^a,\bar{m}^a$ are complex null vectors tangent to the two-spheres 
defined by $u=\text{const}$ and $r=\text{const}$.

While inertial frames of special relativity are related by Poincaré transformations;
Bondi systems are related by the so-called {BMS (Bondi, Metzner and Sachs)} transformations
\begin{eqnarray}
u'&=&K(\zeta,\bar\zeta)(u-\gamma(\zeta,\bar\zeta)),\\
\zeta'&=&\frac{A \zeta + B}{C \zeta + D}, 
\end{eqnarray}
where
\begin{equation}
K(\zeta,\bar\zeta) = \frac{1 + \zeta \bar\zeta}
{(A \zeta + B)(\bar A \bar\zeta + \bar B) + (C \zeta + D)(\bar C \bar\zeta + \bar D)} ,
\end{equation}
$\gamma(\zeta,\bar\zeta)$ an arbitrary real regular function of the angular variables,
and
$(A,B,C,D)$ are complex parameters satisfying $A D - B C =1$. 
The $\gamma$ freedom are known as the supertranslation.

In special relativity, rest frames are
determined by those Cartesian inertial frames 
for which the momentum vector has only the timelike component different from zero;
or in other words, those for which the generator of time translations of the frame is aligned with
the total momentum.

At future null infinity this situation is complicated by the fact that,
although there is a unique definition of {momentum},
there are several definitions of {supermomenta}.
This means that there are several alternative nonequivalent definitions of 
rest frames at $\mathcal{I}^+$.
We nect review some possible definitions of supermomenta.

\subsection{Supermomenta}

Given an arbitrary section $\mathbf{S}$ of $\mathcal{I}^+$, one can choose, without loss of generality,
a Bondi coordinate system $(u,\zeta,\bar\zeta)$, such that $u=u_0=$constant,
determines the section $\mathbf{S}$. Then, the different supermomenta on $\mathbf{S}$ can be expressed in terms of the corresponding integrands as:
\begin{equation} \label{supermomenta}
P_{[X]lm}(\mathbf{S})
=-\frac{1}{\sqrt{4\pi}} 
\int_\mathbf{S} Y_{lm}(\zeta, \bar \zeta) 
\Psi_{[X]}(u=u_0,\zeta,\bar \zeta) 
dS^2 , 
\end{equation}
where $dS^2$ is the surface element of the unit sphere on $\mathbf{S}$,
$Y_{lm}$ are the spherical harmonics, and $[X]$ indicates the type of supermomentum.

Among the different possibilities let us mention the following:
the {Geroch}\cite{Geroch77,Dray84} supermomentum, with integrand
\begin{equation}
 \Psi_{[G]} = \Psi_2^0 + \sigma_0 \dot{\bar \sigma}_0 
+ \frac{1}{2} \left( \eth_0^2 \bar \sigma_0 - \bar\eth_0^2  \sigma_0 \right) ;
\end{equation}
the {Geroch-Winicour}\cite{Geroch81} supermomentum, with integrand\cite{Dain00'}
\begin{equation}
 \Psi_{[GW]} = \Psi_2^0 + \sigma_0 \dot{\bar \sigma}_0 
 - \bar\eth_0^2  \sigma_0  ;
\end{equation}
and the  supermomentum which we have used in the past\cite{Moreschi88} for defining {nice sections},
with integrand
\begin{equation}
 \Psi_{[M]} = \Psi_2^0 + \sigma_0 \dot{\bar \sigma}_0 
+ \eth_0^2 \bar \sigma_0 .
\end{equation}
In all these expressions, dots over quantities $f$ denote Bondi time derivatives, 
 i.e. $\dot{f}=\frac{\partial f}{\partial u}$; 
$\Psi_2^0$  is 
{the leading order part in the asymptotic expansion of} 
the component
\begin{eqnarray}
\Psi_2&=&C_{abcd}\, l^a \, m^b \bar{m}^c n^d \, ,
\end{eqnarray}
of the Weyl tensor $C_{abcd}$, in an expansion in terms of powers of $1/r$ around $\mathcal{I}^+$, i.e.
\begin{equation}
\Psi_2=\frac{\Psi^0_2}{r^3}+\frac{\Psi^1_2}{r^4}+\cdots \; ;
\end{equation}
the scalar $\sigma_0$ is 
{the leading order part in the asymptotic expansion of} 
the shear
\begin{equation}
\sigma=m^am^b\nabla_b l_a ;
\end{equation}
and we use the symbol $\eth_0$ to denote the 
edth operator of the unit sphere.

All these supermomenta have the property that
the first four components of the supermomentum, namely the case $l=0$ and the
three cases $l=1$, determine the Bondi energy-momentum vector. In other words,
\begin{equation} \label{Pbondi} 
\begin{split}
\left(P^{\tt a}\right) = \biggl( &
P_{00},-\frac{1}{\sqrt{6}}(P_{11}-P_{1,-1}), \\
&  \frac{i}{\sqrt{6}}(P_{11}+P_{1,-1}), \frac{1}{\sqrt{3}}P_{10}
\biggr),
\end{split}
\end{equation} 
where $\tt a=0,1,2,3$.

The quantity $\Psi_{[M]}$ has some interesting properties,  
{it is real: $\Psi_{[M]}=\bar\Psi_{[M]}$}, and {also}
 $\dot \Psi_{[M]}=\dot\sigma_0\dot{\bar\sigma}_0.$ 

\subsection{Nice sections}
 
Nice sections provide with a determination of cuts $\mathbf{S}$.
Given an initial section $\mathbf{S}_0$, which it could be thought,
without loss of generality, to coincide with
$u=0$, one can determine any other section, by the supertranslation $\gamma$
that takes from $\mathbf{S}_0$, to   $\mathbf{S}$; which coincides with
the section $u' = u - \gamma = 0$ in a new Bondi coordinate system.

At the section $\mathbf{S}$ we require the supermomentum $P_{[M]lm}(\mathbf{S})$
to have zero spatial components, providing with a geometric notion of rest frame.
In other words, only $P_{[M]00}(\mathbf{S})$ is nonvanishing.
This in general, involves the need to make a Lorentz boost which keeps
$\mathbf{S}$ fixed, but aligns the generator of time translations with the 
total momentum.

It was shown in the past\cite{Moreschi88,Moreschi98} 
that this condition can be cast in the following equation
\begin{equation} \label{nice}
\eth_0^2 \bar \eth_0^2  \gamma =\Psi_{[M]}(\gamma,\zeta,\bar \zeta) 
 +K^3(\gamma,\zeta, \bar \zeta) M(\gamma) \, ,
\end{equation}
where $\gamma$ is the supertranslation which determines the nice section,
$K$ is
the conformal boost factor\cite{Moreschi04},
and
where $M$ is the mass at the section $\mathbf{S}$ given by
\begin{equation}
M=\sqrt{P^{\tt a}P_{\tt a}}.
\end{equation}

Some expected physical 
properties of the nice section equation were proven in references \cite{Moreschi98,Dain00'}, namely:
\begin{itemize} 

\item There exists {a four-parameter family}
of solutions of the nice section equation, for radiating spacetimes.
 
\item Having a nice section
$S_0$, all other nice sections $S_f$ obtained from future timelike translations happen to be to the
future of $S_0$.

\item If the spacetime is stationary, then the nice section equation reduces to 
the good cut equation\cite{Newman66}.
We should emphasize that the good cut equation {only} admits solutions  
in the case of stationary spacetimes, whereas the nice section equation has always solutions.
\end{itemize}

Let us note that if we tried to do a similar construction of nice sections
using instead of the supermomentum $\Psi_{[M]}$, the Geroch supermomentum $\Psi_{[G]}$, 
then one would not be able to obtain equations that determine sections. 
This is due to the fact that
under {BMS} transformations, the expression $\tilde{\eth}^2\tilde{\bar\sigma}_0$ transforms as
\begin{equation}
\begin{split}
\tilde{\eth}^2\tilde{\bar\sigma}_0=&\frac{1}{K^3}\left(\eth^2\bar\sigma_0
-\eth^2\bar\eth^2\gamma\right)+\\
&\frac{1}{K^3}\left[2\eth\gamma\eth\dot{\bar{\sigma_0}}
+\eth^2\gamma\dot{\bar{\sigma_0}}+(\eth\gamma)^2\ddot{\bar{\sigma_0}}\right];
\end{split}
\end{equation}
and therefore the transformation rule for $\Psi_{[G]}$ is
\begin{equation}
\tilde\Psi_{[G]}=\frac{1}{K^3}\left\{\Psi_{[G]}-\frac{1}{2}\left[2\eth\gamma\eth\dot{\bar{\sigma_0}}
+\eth^2\gamma\dot{\bar{\sigma_0}}+(\eth\gamma)^2\ddot{\bar{\sigma_0}}+\text{c.c}\right]\right\}.
\end{equation}
In this way, in the case of a stationary spacetime, we would have no equation for sections, 
since in such situations,
$\Psi_{[G]}$ is supertranslation invariant; in particular, we would not recover the good cuts. 
Also, if we had used the $\Psi_{[GW]}$ supermomentum,  it would not have
the pleasant property of a positive definite time derivative;
which it would complicate the discussion of its properties.

The nice section construction singles out precisely, in an intrinsic way, a Poincaré
structure from the infinite-dimensional BMS group. In particular, given a fixed observational
point $p$ at $\mathcal{I}^+$, there is precisely a three degrees of freedom set of spacelike translations which
generate all the nice sections which contain $p$. In contrast, without this construction there is an
infinite-dimensional family of general sections that contain $p$, one for each supertranslation.
In particular, for stationary spacetimes, the nice sections requirement selects
those whose shear is zero\cite{Moreschi04}.

\section{The Komar angular momentum in axially symmetric spacetimes at null infinity }

Before discussing the charge integral approach to the notion of intrinsic
angular momentum let us review the definition of angular momentum through the Komar integral.
Let $(M,g_{ab})$ be an axially symmetric asymptotically flat spacetime. Therefore, 
it admits a Killing vector $v^a$ with closed orbits, and so it satisfies
\begin{eqnarray}
\nabla_{(b}v_{a)}&=&0.
\end{eqnarray}
Let $\mathbf{S}$ be a sphere, then the Komar integral\cite{Komar59} $K_\mathbf{S}(v)$ is defined by:
\begin{equation}
K_\mathbf{S}(v) =-\frac{1}{16\pi}\int_\mathbf{S}\nabla^a v^b dS_{ab} ;
\end{equation}
where $dS_{ab}$ is the surface element of $\mathbf{S}$.

Let $\mathbf{S}'$ be any other 2-surface, and let $\Sigma$ be a hypersurface which has as boundaries
both, $\mathbf{S}$ and $\mathbf{S}'$.
Then the difference of the Komar integrals at $\mathbf{S}$ and $\mathbf{S}'$ is given by
\begin{equation}
 K_\mathbf{S}(v)  - K_{\mathbf{S}'}(v) = \frac{1}{16\pi}\int_{\Sigma} R^a_{\;b} \; v^b dS_a \; ;
\end{equation}
where $dS_a$ is the volume element of $\Sigma$ and $R^a_{\;b}$ is the Ricci tensor.

This means that: \emph{In vacuum, the Komar integral is conserved.}
It is because of this property that these integrals are so much appreciated.

At future null infinity, 
a rotational Killing vector $v^a$ 
can be thought to be tangent to a sphere $\mathbf{S}$, and
it can be expressed as
\begin{equation}\label{eq:komarf}
v^a = -v_{\bar{m}} \,\hat m^a - v_{m} \, \bar{\hat m}^a
    = -\bar\eth_0\bar{\tilde{a}}\,\hat m^a-\eth_0{\tilde{a}}\,\bar{\hat m}^a ;
\end{equation}
where $\hat m^a$ is 
{the leading order part in the asymptotic expansion of} 
the vector $m^a$,
and we have used the fact that $v_{\bar{m}}$ and $v_{m}$ must be
quantities of spin weight $s=-1$ and $s=1$, respectively; and therefore they can be 
expressed in terms of a spin zero quantity $\tilde{a}$ through the edth operator.

After some computations, which are shown in the Appendix, taking $\tilde a = a$, 
the Komar angular momentum can be written as;
\begin{equation}\label{eq:komar}
K_\mathbf{S}(v) = \frac{1}{{8}\pi}\int_\mathbf{S} \bar\eth_0\bar{\tilde{a}}(\Psi^0_1+\sigma_0\, \eth_0 \bar\sigma_0
)dS^2+\text{c.c.} \; .
\end{equation}

\section{Charge integrals at future null infinity}
It is convenient to approach the concept of physical quantities by the
method of charge integrals.
For example, in electromagnetism, the charge enclosed by a two-surface $\mathbf{S}$ is given by
\begin{equation*}
 Q = -
k
\int_\mathbf{S} {}^*\!F \, ;
\end{equation*}
where $k$ is a constant which depends on the units.

Let us remark that the electromagnetic tensor $F$ can be understood as the {curvature} of the 
{connection} $A$; identified with the potential vector.

In a similar way a charge integral can be obtained from the Riemann curvature tensor.

Given a 2-sphere  $\mathbf{S}$, we will work with the 
charge integral of the Riemann tensor:
\begin{equation}
  \label{eq:chargeinte}
  Q_\mathbf{S} =\int_\mathbf{S}C 
\end{equation}
where the 2-form $C_{ab}$ is given by
\begin{equation}
  \label{eq:chargeint}
  C_{ab} \equiv R_{ab}^{*\;\;cd}\; w_{cd} ,
\end{equation}
with $R_{ab}^{*\;\;cd}=\frac{1}{2}R_{abef}\epsilon^{efcd}$,
the right dual of the Riemann tensor, 
and $w_{ab}$ a 2-form which will be determined next. For motivations of these type of
charge integrals see\cite{Penrose82, Moreschi86,Goldberg90}.

At this point, one might wonder why we use the right dual of the Riemann tensor instead of the 
left dual, the answer is that using the right dual
we can directly relate the exterior derivative of the form $C_{ab}$ with the 
Einstein tensor as we show below.

Let $\Sigma$ be a spacelike hypersurface in the interior of the spacetime but which asymptotically reaches future null infinity in such a way that in the conformally completed
spacetime, $\Sigma$ can be extended to $\mathcal{I}^+$ with boundary $\mathbf{S}$. 
Then, by using Stokes' theorem
the charge integral on $\mathbf{S}$ can be expressed as an integral on $\Sigma$, namely
\begin{equation}
  \label{eq:qonsigma}
    Q_\mathbf{S} =\int_\mathbf{S}C = \int_{\Sigma} dC .
\end{equation}

The exterior derivative of $C$ can be expressed\cite{Moreschi86} by
\begin{equation}
  \label{eq:dC}
  dC_{abc} =\frac{1}{3} \epsilon _{abcd} \;^{*}R^{*defg} \;\nabla _{e} w_{fg}
.
\end{equation}
As it was said before, an important property of the double dual
of the Riemann tensor is that its trace gives the Einstein tensor,
namely:
\begin{equation}
 {}^{*}R_{abcd}^{*} g^{bd} =G_{ac} =R_{ac} -\frac{1}{2} g_{ac} R \; .
\end{equation}
Therefore the previous equation can be written as
\begin{equation}\label{eq:dC2}
\begin{split}
dC_{abc}  
= & \frac{1}{3} \epsilon _{abcd} \left( -\frac{2}{3}G^{dg} v_{g} + {}^{*}R^{*defg}
\ T_{efg} \right) ;
\end{split}
\end{equation}
where $T_{abc}$ is the traceless part of $\nabla_a w_{bc}$ and
$v_c$ its trace; that is:
\begin{equation}
  \label{eq:nablaw}
  \nabla _{a} w_{bc} = T_{abc} +\frac{1}{3} g_{ab} v_{c} -\frac{1}{3}
g_{ac} v_{b} ;
\end{equation}
with
\begin{equation}\label{eq:divw0}
 \nabla_a w^{ab}=v^b \; .
\end{equation}

Let us study, for a moment, this expression from the point of view of linearized gravity.
Suppose that the metric is expressed as $g=\eta+h$, in terms of a flat
background metric $\eta$.
Then, from equations (\ref{eq:dC2}) and (\ref{eq:divw0}) one observes that if the vector $v^a$
is a Killing vector of the metric $\eta_{ab}$ and $T_{abc}$
is $O(h)$, then the charge integral will give the
conserved quantities in the context of linearized gravity. 
It is clear that one can always find such a $w$.
Then, this analysis ensures that this {charge integrals admit the appropriate
physical interpretations in the linearized gravity regime}, and in particular that they 
do not suffer from the factor of two anomaly\cite{Winicour80}.

Another property of the double dual of the Riemann tensor is
the one associated with the Bianchi identities, namely 
${}^*R^{*d[efg]} = 0$, from which one can prove\cite{Moreschi04} the relation
\begin{equation}
  {}^*R^{*defg} \, T_{efg} 
=    \frac{2}{3}\; {}^*R^{*defg} \left(T_{(ef)g} - T_{(eg)f} \right) \; .
\end{equation}

This expression can be written
in more simple form using spinorial
notation. $T_{abc}$
can be expressed\cite{Moreschi04} as,
\begin{equation}
  \label{eq:Tofw}
\frac{2}{3} \;\left( T_{(ef)g} -T_{(eg)f}
\right) =\nabla _{E' (E} w_{FG)} \;\epsilon _{F' G' } + {\tt c.c.} \;;
\end{equation}
where ${\tt c.c.}$ means complex conjugate,
and we have made the standard abuse of notation identifying 
the vectorial abstract indices with spinorial abstract indices
using the rule $e = E E'$.

From these considerations, the most natural conditions on $w$ are 
to stay as close as possible to the following conditions
\begin{equation}
  \label{eq:divw}
  -\nabla _{A} ^{\;\;B' }\; w^{AB} + {\tt c.c.}= v^{BB'} \; ,
\end{equation}
and 
\begin{equation}
  \label{eq:symw}
  \nabla _{E' (E} \; w_{FG)} =0;
\end{equation}
where the vector $v^{BB'} $ is a generator of asymptotic symmetries.

In general an asymptotic symmetry $v^a$ can be expressed by 
its components, in terms of a null tetrad frame
\begin{equation}
  \label{eq:vasym}
  v^a = v_n \, \ell^a -v_{\bar m}\,  m^a -v_m \, \bar m^a + v_\ell \, n^a .
\end{equation}

Since the asymptotic symmetries are tangent to $\mathcal{I}^+$, the
tetrad components have the following behavior
\begin{eqnarray}
  \label{eq:vn}
  v_n &=& r v^0_n + v^1_n + O(\frac{1}{r}) ,\\
  \label{eq:vm}
  v_m &=& r v^0_m + v^1_m + O(\frac{1}{r}) ,\\
  \label{eq:vl}
  v_\ell &=&  v^0_\ell + \frac{v^1_\ell}{r} + O(\frac{1}{r^2}) .
\end{eqnarray}

The 
{leading order part in the asymptotic expansion} 
in a Bondi system of an asymptotic symmetry
is given by
\begin{equation}
  \label{eq:vm0}
  v^0_{m} =\eth_0 a ,
\end{equation}
\begin{equation}
  \label{eq:vn0}
  v^0_{n} =\frac{1}{2} \left(\eth_0 v^0_{\bar{m} } 
    + \bar\eth_0 v^0_{m} \right) =\frac{1}{2}\eth_0 \bar\eth_0 (a+\bar{a} ) ,
\end{equation}
\begin{equation}
  \label{eq:vl0}
  v^0_{l} =\chi (\zeta ,\bar{\zeta } )-u\frac{1}{2} \eth_0 \bar\eth_0(a+\bar{a} ) ;
\end{equation}
where $\chi $ and $a$ are functions on the sphere with spin weight 0,
satisfying $\chi=\bar\chi$, 
 $\dot{a} =0$ and
 $\eth_0^{2} a=0$.

Relation (\ref{eq:divw})  at $\mathcal{I}^+$ can 
be expressed  in terms of the spinorial components of a regular dyad
\begin{equation}
  \label{eq:wcomp}
w^{AB} =w_{0} \;\hat{\iota}^{A} \hat{\iota}^{B} - w_{1} 
\left(\hat{o}^{A} \hat{\iota}^{B} + \hat{\iota}^{A} \hat{o}^{B} \right)
+ w_{2} \;\hat{o}^{A} \hat{o}^{B}   ,
\end{equation}
by
\begin{eqnarray}
  \label{eq:w2}
  w_{2} &=&-\frac{1}{3} v^0_{\bar{m} } ,\\
  \label{eq:w1}
  w_{1} +\bar{w} _{1} &=&-\frac{1}{3} v^0_{\ell} ,\\
  \label{eq:dotw1}
  \dot{w} _{1} +\dot{\bar{w} } _{1} &=&-\frac{1}{2} \left(\eth_0 w_{2} 
+\bar\eth_0 \bar{w}_{2} \right) ;
\end{eqnarray}
while condition (\ref{eq:symw}) at $\mathcal{I}^+$ becomes
\begin{eqnarray}
  \label{eq:ethw2}
  \bar\eth_0   w_{2} &=&0 ,\\
  \label{eq:dotw2}
  \dot{w} _{2} &=&0 ,\\
  \label{eq:ethw0}
  \eth_0 w_{0} &=&-2\;\sigma_0 \;w_{1} ,\\
  \label{eq:dotw1b}
  \dot{w} _{1} &=&-\frac{1}{2} \eth_0 w_{2} ,\\
  \label{eq:dotw0}
  \frac{1}{2} \dot{w} _{0} + &\eth_0 w_{1}& +\sigma_0 \;w_{2} =0 .
\end{eqnarray}

It was shown in reference \cite{Moreschi04} that the charge integral at future
null infinity can be expressed as:
\begin{eqnarray}
  \label{eq:chargescri2}
    Q_\mathbf{S}(w) &=& 4 \int \left[
-  w_2 \Psi_1^0  + 
  2  w_1 (\Psi_2^0 + \sigma_0 \dot{\bar \sigma}_0)
\right] dS^2\nonumber\\
&& + {\tt c.c.};
\end{eqnarray}
where $\Psi_1^0$, is 
{the leading order part in the asymptotic expansion of} 
the 
{tetrad component of the Weyl tensor} 
\begin{eqnarray}
\Psi_1&=&C_{abcd}l^a n^b m^c l^d .
\end{eqnarray}


\section{Charge integrals in stationary spacetimes and physical quantities}


For the case of stationary spacetimes one can solve the set of equations
(\ref{eq:w2})-(\ref{eq:dotw0}) with solution
\begin{align}
w_{2} &=-\frac{1}{3} \bar\eth_0 \bar a  , \label{eq:wsolut1} \\
w_{1} &=w_{1}^{00} ( \zeta ,\bar{\zeta },\sigma_0  )+
           \frac{1}{6}\,u\, \eth_0 \bar\eth_0 \bar a  , \label{eq:wsolut2}\\
w_{0} &=w_{0}^{00} (\zeta ,\bar{\zeta }, \sigma_0 ) +u\left( -2 \eth_0 w_{1}^{00} 
           + \frac{2}{3} \sigma_0  \bar\eth_0 \bar a 
         \right) \nonumber \\ 
& \quad - \frac{1}{6}\, u^2 \, \eth_0^{2} \bar\eth_0 \bar a  ; \label{eq:wsolut3}
\end{align}
where $\bar\eth_0^2 \bar a =0$,
and $w_{1}^{00}$ 
and $w_{0}^{00}$ are spin weight 0 and 1 functions 
respectively that solve the equations
\begin{equation}
 \label{eq:w100}
\eth_0 ^{2} w_{1}^{00} =\frac{1}{3}\eth_0 \sigma_0 \;\bar{\eth_0}\bar{a} +
\frac{1}{2} \sigma_0\;\eth_0 \bar{\eth_0}\bar{a} =
-\eth_0\sigma_0 \;w_{2} -\frac{3}{2} \sigma_0 \;\eth_0 w_{2} \; ,
\end{equation}
and
\begin{equation}
 \label{eq:w000}
\eth_0  w_{0}^{00} =-2\sigma_0 w_{1}^{00} .
\end{equation}

Let us note that if one uses the potential $\delta$ of the shear satisfying
\begin{equation}
 \label{eq:alfasigma}
  \sigma_0 = \eth_0^2 \delta ;
\end{equation}
then, the component $w_1$ can be expressed by
\begin{equation}
 \label{eq:w1alfa}
 w_1 = b + \frac{1}{3} \eth_0 \delta \bar\eth_0 \bar a 
+ \frac{1}{6} (u - \delta) \eth_0 \bar\eth_0 \bar a ;
\end{equation}
where the spin 0 quantity $b$ satisfies $ \dot b =0$ and $\eth_0^2 b = 0$.

This procedure provides with
a two-form { $w_{AB}^{0}$ given by equations (\ref{eq:wsolut1})-(\ref{eq:wsolut3})}
 with the functional dependence on $u=\gamma$ given by:
\begin{equation}
 \label{eq:w0}
  w_{AB}^{0}(\gamma) =w_{AB}^{0} \left( 
u=\gamma,\zeta ,\bar{\zeta } 
;
\sigma_0(\zeta ,\bar{\zeta } ),a,b
\right) \, ;
\end{equation}
where we stress the dependence on $\sigma_0$.

Let us observe that $a$ involves 6 real constants associated with the
Lorentz rotations, and that since in this case $\Psi_2^0$ is a real quantity, $b$
contributes to the charge integral $Q_\mathbf{S}(w)$ with four other real constants
associated with translations.

The first term in the integrand of equation (\ref{eq:chargescri2}) includes
the Weyl component $\Psi_1$ which is known to describe the angular
momentum in the Kerr geometry. In the second term
we recognize the component $\Psi_2^0$ which determines the
supermomentum $\Psi_{[M]}$ for this particular
stationary case.

Let us recall that
in special relativity, angular momentum and intrinsic angular momentum
are related by expressions of the form
$J^{\tt a  b}=S^{\tt a b}+R^{\tt a} \, P^{\tt b} - P^{\tt a}\, R^{\tt b}$.
Then, given a rest reference
frame in Minkowski spacetime, one needs to use the spacelike translation
freedom $(R^{\tt b})$ 
in order to single out
the center-of-mass reference frame. In the center-of-mass frame
one has $ J^{\tt a  b}=S^{\tt a b}$. 
Also, since the intrinsic
angular momentum satisfies $S^{\tt ab} P_{\tt b}=0$ 
(frequently referred to as Dixon condition\cite{Dixon:1970zza}), 
one can characterize
the center-of-mass frame as that rest frame for which 
$J^{0i}=0$. 
It can be seen that the condition we need to impose on the section $\mathbf{S}$ in the charge integral case is
that it must be the nice section satisfying:
\begin{equation}
  \label{eq:acondit}
  Q_\mathbf{S}(a)=0 \qquad \text{for all}\quad a=\bar a ;
\end{equation}
where it is understood that one takes $b=0$ in this equation.
The quantity $a$ is, in principle, complex; so this condition
makes use of precisely of 3-degrees of freedom, which are
associated with spacelike translations.

This is the appropriate condition which leaves a one-dimensional family of
nice sections $\mathbf{S}$ that can legitimately be called 
{\em center-of-mass frames}.
In particular, 
we can see that the other center-of-mass frames  are generated by time translations,
from an original one,
in the nice section construction.
Using these frames $\mathbf{S}_{\tt cm}$, the intrinsic angular momentum 
$\mathtt{s}$ is defined through
\begin{equation}
  \label{eq:instrin1}
  \mathtt{s} = \frac{3}{{32} \pi} Q_{\mathbf{S}_{\tt cm}}(w) ;
\end{equation}
where to determine $w$ one chooses $a= -\bar a$ and $b=0$.
Note also that the same charge integral can be used to calculate the 
Bondi momentum $\mathtt{p}$ given by 
\begin{equation}
  \label{eq:moment2}
  \mathtt{p} = \frac{3}{{32} \pi} Q_{\mathbf{S}_{\tt cm}}(w) ;
\end{equation}
where in this case one
takes $a=0$ and $b\neq0$.

The previous prescription singles out the center-of-mass frame for stationary
spacetimes and a Poincaré subgroup of BMS generators.

At this point, it is probably worthwhile to mention that some authors have considered
an alternative to Dixon condition, known as Mathisson condition\cite{Mathisson37};
 which requires $J^{ab}U_b=0$, with $U^b$ 
the four-velocity vector
of the worldline that they associate to the center-of-mass. 
But this condition has the difficulty that it does not prescribe a 
unique worldline;
in fact the solution depends on the choice of an initial $U^b$; 
which gives rise to helical motion for free spinning particles, instead
of the geodesic motion. 
Some authors had given physical meaning to these curves\cite{Moller49}
and recent articles discuss the range of validity (see for example \cite{Costa:2011zn}).
We do not make use of this alternative condition, because we require a definition of 
center-of-mass and intrinsic angular momentum to be constructed from intrinsic physical
quantities avoiding arbitrary choices.



\section{Intrinsic angular momentum for radiating spacetimes which agrees with the Komar integral}

We show in detail how to obtain the intrinsic angular momentum for the general radiating case.

As before, we define the rest frame sections as those for which:
$P_{[M]lm}(\mathbf{S})=0, \;\;\forall \;l\geq 1.$

Let us consider a point along a particular generator of $\mathcal{I}^+$, denoted by $p(\tau)$,
with $\tau$ a monotonically increasing time parameter. The set of nice sections form a four-parameter $(T,\vec R)$
family  that we now label $\mathbf{S}_{(T,\vec R)}$; 
where  $(T,\vec R)$ can be identified with a {translation among nice sections}.
Then, for a given fixed $\tau$, one has a 3-parameter 
family of nice sections 
$\mathbf{S}_{(T,\vec R)}$ which contains the point $p(\tau)$.

Given one of these nice sections $\mathbf{S}_{(T,\vec R)}$, we can always identify it
with the condition $u=\gamma_\mathbf{S}$, where $\gamma_\mathbf{S}(\zeta,\bar\zeta)$ is the supertranslation
that defines the corresponding section. 

At this point it is important to emphasize that equations (\ref{eq:ethw2})-(\ref{eq:dotw0}),
in general, do not have solutions  in a radiating spacetime.
In spite of that, we can propose a prescription which defines a 2-form suitable for
our construction.

On $\mathbf{S}_{(T,\vec R)}$ we define the 2-form 
{$w_{\mathbf{S}_{(T,\vec R)}}$}
{as the solution of the stationary problem ((\ref{eq:wsolut1})-(\ref{eq:wsolut3}))} 
where the radiation data is taken as
$\sigma_{\mathbf{S}_{(T,\vec R)}}(u,\zeta,\bar\zeta) =  \sigma_0(\gamma_\mathbf{S},\zeta,\bar\zeta)$.

Then, using the identity 
\begin{equation}
  \label{eq:w1ethsigma}
\int\limits_{\mathbf{S}}\left( w_{1}\;\eth_0^2 \bar{\sigma_0 } \right) \ dS^{2}  
=
\int\limits_{\mathbf{S}}\left( \sigma_0 \, \eth_0\bar{\sigma}_0  
  +\frac{1}{2} \eth_0 \left(
\sigma_0 \bar{\sigma}_0  \right) \ \right) w_{2} \ dS^{2} ;
\end{equation}
we see that the charge integral (\ref{eq:chargescri2}) can be expressed more generally as
\begin{equation}
  \label{eq:charge2}
\begin{split}
Q_\mathbf{S}(w)
 =4 
\int_\mathbf{S}\Bigl\{ & \, -w_{2} \left[ \Psi_{1}^{0} 
-{\alpha}\left(\sigma_0 \eth_0\bar{\sigma_0 } +
\frac{1}{2}\eth_0\left( \sigma_0 \bar{\sigma}_0  \right)\right)\right] 
+ \\
& 2 \,w_{1} \left[ \Psi _{2}^{0} +\sigma_0 \dot{\bar{\sigma_0 } } 
-\frac{{\alpha}}{2}\eth_0^{2} \bar{\sigma_0 } \right] \Bigr\} + {\tt c.c.} ,
\end{split}
\end{equation}
with {$\alpha$} a constant.
Let us note that using the potential $w_2$,
the Komar angular momentum (\ref{eq:komar}) can be written as;
\begin{equation}\label{eq:komar2}
K_\mathbf{S}(v) = \frac{3}{{8}\pi}\int_\mathbf{S} -w_2(\Psi^0_1+\sigma_0\, \eth_0 \bar\sigma_0
)dS^2+\text{c.c.} \; .
\end{equation}  
Therefore, in order to recover the Komar expression for the angular momentum, in the case 
of axisymmetric spacetimes, we must set $\alpha=-1$.

In this way,
the expression for the charge integral,
that it can be used to calculate angular momentum, momentum or supermomentum,
is:
\begin{equation}
  \label{eq:charge3}
\begin{split}
Q_\mathbf{S}(w)
 =4 
\int_\mathbf{S}\Bigl\{ & \, -w_{2} \left[ \Psi_{1}^{0} 
+\sigma_0 \eth_0\bar{\sigma_0 } +
\frac{1}{2}\eth_0\left( \sigma_0 \bar{\sigma}_0  \right)
\right] \\
&+ \, 2 \,w_{1} \left( \Psi_{[M]}  -\frac{1}{2}
\eth_0^{2} \bar{\sigma_0 } \right) \Bigr\} + {\tt c.c.} .
\end{split}
\end{equation}

It is interesting to note that the first integrand factor coincides with that obtained 
by Winicour\cite{Winicour80} (except for relative signs, due 
to difference in conventions).

Let us note then, that the first term takes the form of the Komar angular momentum;
since one can check that 
when $3 w_2$ is the component of a
rotational Killing vector, the term 
$\eth_0\left( \sigma_0 \bar{\sigma}_0  \right)$, does not contribute. 
Also, let us observe that in the case of a stationary spacetime, 
it reduces to the result of the previous section, since the center-of-mass sections
coincide with the sections with $\sigma_0 = 0$.

In addition, in order to compute the intrinsic angular momentum,
for each choice of $a$ we take $b(a)$ which satisfies:
\begin{equation}\label{eq:bdea}
\begin{split} 
b  =& \frac{1}{48 \pi M}
\int_{\mathbf{S}}
\left(
- \eth_0\delta\bar\eth_0\bar{a}
+ 
\frac{1}{2}
\delta\eth_0\bar\eth_0\bar a\right)\eth_0^2\bar\eth_0^2 \bar\delta \, dS^2\\
&+ \text{c.c.};
\end{split}
\end{equation}
where $\delta$ is the complex potential for the shear defined by
$$\sigma=\eth_0^2\delta \; .$$
This choice of $b$ is made so that the second term in (\ref{eq:charge3}) does
not contribute and so the complete charge integral coincides with the Komar integral.

Similarly as it was done before,
in order to single out the center-of-mass section $\mathbf{S}_{\tt cm}$ from the 3-parameter
family of nice sections which contain the point $p(\tau)$, we 
demand 
\begin{equation}
  \label{eq:acondit2}
  Q_{\mathbf{S}_{\tt cm}}(a)=0 \qquad \text{for all}\quad a=\bar a .
\end{equation}

Using these center-of-mass frames $\mathbf{S}_{\tt cm}$, the intrinsic angular momentum 
$\mathbf{s}$ is defined through
\begin{equation}
  \label{eq:instrin}
  \mathbf{s} = \frac{3}{{32} \pi} Q_{\mathbf{S}_{\tt cm}}(w) ;
\end{equation}
where as before, in order to pick up the intrinsic angular momentum, one
takes $a=-\bar a$.


In this approach, the observer (us) is located at future null infinity, confined
to a particular generator of it. 
Then, for each retarded time, our construction singles out a unique
center-of-mass section, where the calculation of the intrinsic angular momentum is
carried out. Thus, if one wants to compare the spin at two different times,
one has to repeat the construction of the center-of-mass section at the
second reference time.
Let us also note that since between two center-of-mass sections we have at our
disposal the one parameter family of center-of-mass sections among them,
we also have the two-form $C_{ab}$, on this region, and therefore,
the difference of the spins can be expressed as a flux law,
using Stokes' theorem.


\section{Final comments}

We have shown in detail how the problem of supertranlations, in defining angular momentum,
 can be circumvented with the help of the so called nice sections and the charge integrals.

The comparison of the present approach based on charge integrals of the Riemann tensor
with traditional approaches has been done in reference \cite{Moreschi04}.
Among the recent contributions on the subject we comment on the ingenious work
of reference \cite{Szabados08} based in Dirac eigenspinors.
The application of the so called spectral angular momentum
to our example\cite{Gallo09} of a supertranslated boosted section in Schwarzschild
spacetime, captures the nonzero value of the orbital angular momentum.
In contrast, in our approach, the intrinsic angular momentum for 
any center-of-mass section, gives the expected zero value.

As it was mentioned in the introduction we emphasize again that our work
tackles the definition of intrinsic angular momentum as opposed to 
just total angular momentum; as discussed by most other works.
It is the notion of intrinsic angular momentum that is relevant to
the study of astrophysical systems and in particular to the
problem of balance of gravitational radiation.

Another point that it is worthwhile to remark, is that the definition 
presented here, satisfies the property that: in a spacetime with
three stages -an axisymmetric one, a nonaxisymmetric one,
and a third with a different Killing symmetry- it gives the 
expected values at the first and third stages.

In summary, we have presented a definition of intrinsic angular momentum
which is free from supertranslation dependence. It can be applicable to
a general radiating spacetime, and it agrees, for the case of axial symmetry
with the Komar integral.

To our knowledge this is the only definition of intrinsic angular momentum
with these properties.

\subsubsection*{Acknowledgements}
We are grateful to László Szabados for deep illuminating discussions and 
for kind hospitality at
the Wigner Research Centre for Physics, KFKI, Budapest.

We would also like to acknowledge the valuable comments and criticism from
anonymous referees, which help us to considerably improve the presentation
of our work.

We acknowledge financial support from CONICET and SeCyT-UNC.

\setcounter{equation}{0}
\renewcommand{\theequation}{\thesubsection A\arabic{equation}}
{\section*{Appendix} 
\subsection*{The Komar expression for angular momentum}}


In an axisymmetric spacetime there exist a Killing vector field $v^a$ 
associated to the axial symmetry.
If the spacetime is also asymptotically flat, then we the Komar definition of angular momentum is
\begin{equation}\label{eq:jkomar}
K_\mathbf{S}(v)=-\frac{1}{16\pi}\int_\mathbf{S} \nabla^bv^{a}dS_{ab},
\end{equation}
where $dS_{ab}$ is the surface element of a two-sphere $\mathbf{S}$ defined as a cut of $\mathcal{I}^+$.
An operational way to explicitly write this integral is to consider a Bondi 
system $(u,\zeta,\bar\zeta)$ such that
we extend $\mathbf{S}$ to the interior of the spacetime 
along null geodesics with tangent vector $l^a$ such that they are
orthogonal to $\mathbf{S}$ at $\mathcal{I}^+$. 
This construction generates a null surface given by $u=\text{constant}$, with $l_a=(du)_a$. 
If we also define a affine 
parameter $r$ along the null geodesics $l^a$, then the two-surfaces $S_{u,r}$ 
defined by $r=\text{constant}$ on $u=\text{constant}$ 
will be two-spheres.
We can then complete $l^a$ to a null tetrad 
$\left\{l^a,n^a,m^a,\bar{m}^a\right\}$, by doing $m^a$, and $\bar{m}^a$  
tangent to the two-spheres $S_{u,r}$. 
Then, the Komar angular momentum (\ref{eq:jkomar}), can be reexpressed as: 
\begin{equation}\label{eq:Jlimit}
K_\mathbf{S}(v)=-\frac{1}{8\pi}\lim_{r\rightarrow\infty}\int_\mathbf{S} \nabla^b v^{a}l_{[a}n_{b]}d\tilde{S}^2;
\end{equation}
where we used $dS_{ab}=2l_{[a}n_{b]}d\tilde{S}^2$, with $d\tilde{S}^2$ the surface-area element of the spheres $S_{u,r}.$ 
Now, the Killing vector $v^a$ is, by construction, tangent to the two-spheres $S_{u,r}$; then,
it must be expressed as
\begin{equation}\label{eq:komarf2}
v^a=-v_{\bar{m}}m^a-v_{m}\bar{m}^a=-\bar{\eth}\bar{\tilde{a}}\,m^a-\eth{\tilde{a}}\,\bar{m}^a.
\end{equation}
where in the last equality we used the fact that $v_{\bar{m}}$ and $v_{m}$ are
quantities of spin weight $s=-1$ and $s=1$, respectively, and therefore can be 
written in terms of a spin-zero quantity  $\tilde{a}$ through the edth operator.
Then, by proyecting the Killing equation $\nabla_{(a}v_{b)}=0$ in the 
direction of $m^a\bar{m}^b$, we get 
\begin{equation}
\eth\bar\eth\left(\tilde{a}+\bar{\tilde{a}}\right)=0,
\end{equation}
This equation implies that $\tilde{a}+\bar{\tilde{a}}=\text{constant}$; 
therefore, 
without loss of generality we can take  $\tilde{a}+\bar{\tilde{a}}=0$.
On the other hand, the integrand of (\ref{eq:Jlimit}) reads
\begin{equation}
\begin{split}
\nabla^b v^{a}l_{[a}n_{b]}=&-\nabla^b\left(\bar\eth\bar{\tilde{a}}\,m^a
+\eth{\tilde{a}}\,\bar{m}^a\right)l_{[a}n_{b]}\\
=&\;\;\eth{\tilde{a}}\bar\tau\,
+\bar\eth\bar{\tilde{a}}\,\tau,
\end{split}
\end{equation}
with 
\begin{equation}
\tau=m^an^b\nabla_b l^a.
\end{equation}
This means that in order to compute the angular momentum defined 
in (\ref{eq:Jlimit}), we need to know the $O(r^{-2})$ of 
$\eth\bar{\tilde{a}}\,\bar\tau$. 
It can be shown, from the Killing equations, that the term $\bar\eth\bar{\tilde{a}}$ 
has an asymptotic expansion as
\begin{equation}
\bar\eth\bar{\tilde{a}}=v^0_{\bar{m}}r+v^1_{\bar{m}}+O(r^{-1}),
\end{equation}
with $v^0_{\bar{m}}$ and $v^1_{\bar{m}}$ functions of $(u,\zeta,\bar\zeta)$. 
From the integration of the Killing equation
\begin{equation}
\text{\Thorn}v_{{m}}=-\sigma v_{\bar{m}}-\bar\rho v_{{m}};\label{eq:kil1}
\end{equation}
where \Thorn\, is the thorn operator in the GHP\cite{Geroch73} notation,
and taking into account the expansion in powers of $r$ of $\rho$ and $\sigma$,
\begin{eqnarray}
\rho &=&-\frac{1}{r}+O(r^{-3}),\\
\sigma &=&\frac{\sigma_0}{r^{2}}+O(r^{-4});
\end{eqnarray}
it also follows that
\begin{equation}
v^1_{\bar{m}}=\bar\sigma_0v^0_{{m}}.
\end{equation}
Let also note, that in terms of a regular tetrad, the 
Killing vector on $\mathcal{I}^+$ is given by:
\begin{equation}
v^a|_{\mathcal{I}^+}=-{v^0_{\bar{m}}}\hat{m}^a-v^0_{{m}}\hat{\bar{m}}^a.
\end{equation}
On the other hand, because $v^0_{\bar{m}}$ is a quantity 
of spin-weight $s=-1$, it can be expressed in terms of a 
spin-weight $0$ quantity $\bar{a}$ as
\begin{equation}
v^0_{\bar{m}}=\bar\eth_0{\bar{a}}\; .
\end{equation}

In the same way, it can be shown that $\tau$ has an asymptotic expansion as
\begin{equation}
\tau=\bar\eth_0\sigma_0r^{-2}-(\Psi^0_1+2\sigma_0\eth_0\bar\sigma_0)r^{-3}+O(r^{-4}).
\end{equation}
By expanding the product $\bar\eth\bar{\tilde{a}}\,\tau$ up to order $O(r^{-2})$, we obtain
\begin{equation}
\begin{split}
\bar\eth\bar{\tilde{a}}\,\tau
=&\bar\eth_0\sigma_0 v^0_{\bar{m}}r^{-1}\\
&+\left\{-\left[\Psi^0_1+2\sigma_0\eth_0\bar\sigma_0\right]v^0_{\bar{m}}+
\bar\sigma_0\bar\eth_0\sigma_0v^0_{{m}}\right\}r^{-2}\\
&+O(r^{-3}).
\end{split}
\end{equation}
Let us note from the first line of 
the previous equation, that not only does the leading order 
of the Killing vector make a contribution to the integral 
(given by $v^0_{\bar{m}}$) but it also contributes the term given by $v^1_{\bar{m}}$.
Let us also note, that the first term of this expression can be written as
\begin{equation}
\begin{split}
\bar\eth_0\sigma_0 v^0_{\bar{m}}r^{-1}=&-\bar\eth_0\sigma_0\bar\eth_0{\bar{a}}r^{-1}\\
=&-\left[\bar\eth_0\left(\sigma_0\bar\eth_0{\bar{a}}\right)
-\sigma_0\bar\eth^2_0{\bar{a}}\right]r^{-1}\\
=&-\bar\eth_0\left(\sigma_0\bar\eth_0{\bar{a}}\right)r^{-1};
\end{split}
\end{equation}
where we used the fact that $\bar\eth^2_0\bar{a}=0$, which follows 
from projecting the Killing equation $\nabla_{(a}v_{b)}=0$ in the 
direction of $m^a{m}^b$.
Therefore by adding its complex conjugate we obtain,
\begin{equation}
\begin{split}
\eth{\tilde{a}}\,\bar\tau+\bar\eth\bar{\tilde{a}}\tau
=&-\left[\bar\eth_0\left(\sigma_0\bar\eth_0{\bar{a}}\right)
+\eth_0\left(\bar\sigma_0\eth_0{{a}}\right)\right]r^{-1}\\
&+\left[-(\Psi^0_1+\sigma_0\eth_0\bar\sigma_0)\bar\eth_0{\bar{a}}
+\text{c.c}\right]r^{-2}
+O(r^{-3}).
\end{split}
\end{equation}
Then, the angular momentum reads;
\begin{equation}
K_\mathbf{S}(v)=\frac{1}{8\pi}\int_\mathbf{S}(\Psi^0_1+\sigma_0\eth_0\bar\sigma_0)
\bar\eth_0{\bar{a}}dS^2+\text{c.c}.
\end{equation}
with $dS^2$ the surface element of a unit two-sphere.
This expression can be written in terms of
\begin{equation}
w_2=-\frac{1}{3}\bar\eth_0\bar{a};
\end{equation} 
as
\begin{equation}
K_\mathbf{S}(v)=-\frac{3}{8\pi}\int_\mathbf{S} w_2(\Psi^0_1+\sigma_0\eth_0\bar\sigma_0
)dS^2+\text{c.c}.
\end{equation}


\begingroup\raggedright\endgroup

\end{document}